\documentclass{webofc}
\usepackage[varg]{txfonts}   % Web of Conferences font
\usepackage{hyperref}
\usepackage{url}
\usepackage{lineno}%%%%%%%%%%%%%%%%%%%%%%%%%%%%%%%%%%%%%%%%%%%%%%%%%%%%%%%%%%%%%%%%%%%%%%%%%%%%%
\hypersetup{colorlinks=true,citecolor=blue,urlcolor=blue,linkcolor=blue}

\newcommand{\pp}           {pp\xspace}

\newcommand{\pPb}          {\mbox{p--Pb}\xspace}

\newcommand{\s}            {\ensuremath{\sqrt{s}}\xspace}

\newcommand{\pt}           {\ensuremath{p_{\rm T}}\xspace}

\newcommand{\jpt}           {\ensuremath{p_{\rm T}^{\rm ch\> jet}}\xspace}

\newcommand{\rl}           {\ensuremath{R_{\rm L}}\xspace}
\newcommand{\ptrl}         {\ensuremath{\langle p_{\rm T}^{\rm ch\> jet}\rangle R_{\rm L}}\xspace}

\newcommand{\EEC}          {\ensuremath{\Sigma_{\rm EEC}}\xspace}

\newcommand{\nineH}        {$\sqrt{s}~=~0.9$~Te\kern-.1emV\xspace}
\newcommand{\seven}        {$\sqrt{s}~=~7$~Te\kern-.1emV\xspace}
\newcommand{\twoH}         {$\sqrt{s}~=~0.2$~Te\kern-.1emV\xspace}
\newcommand{\twosevensix}  {$\sqrt{s}~=~2.76$~Te\kern-.1emV\xspace}
\newcommand{\five}         {$\sqrt{s}~=~5.02$~Te\kern-.1emV\xspace}
\newcommand{\twosevensixnn}{$\sqrt{s_{\mathrm{NN}}}~=~2.76$~Te\kern-.1emV\xspace}
\newcommand{\fivenn}       {$\sqrt{s_{\mathrm{NN}}}~=~5.02$~Te\kern-.1emV\xspace}

\newcommand{\GeVc}         {Ge\kern-.1emV/$c$\xspace}
\newcommand{\MeVc}         {Me\kern-.1emV/$c$\xspace}
\newcommand{\TeV}          {Te\kern-.1emV\xspace}
\newcommand{\GeV}          {Ge\kern-.1emV\xspace}
\newcommand{\MeV}          {Me\kern-.1emV\xspace}
\newcommand{\GeVmass}      {Ge\kern-.2emV/$c^2$\xspace}
\newcommand{\MeVmass}      {Me\kern-.2emV/$c^2$\xspace}

\begin{document}
\title{Energy-energy correlators in jets across collision systems and parton flavors}

\author{
\firstname{Anjali} \lastname{Nambrath}\inst{1}\fnsep\thanks{\email{nambrath@berkeley.edu}} 
\firstname{} \lastname{on behalf of the ALICE collaboration} \inst{}
}

\institute{University of California, Berkeley}

\abstract{
The two-point energy-energy correlator (EEC) is a novel jet substructure observable probing the correlation of energy flow within jets. In these proceedings, three EEC measurements performed by the ALICE Collaboration are reported. First is a finalized measurement in proton-proton collisions, where the angular dependence of the EEC cross-section shows a separation of the perturbative and non-perturbative regimes. The first EEC measurement is reported in heavy-flavor jets, tagged via a fully-reconstructed D$^0$ meson. Here, comparison to inclusive \pp jets provides insights into the flavor dynamics of QCD fragmentation and hadronization. Finally, the first measurement of EECs in \pPb collisions is presented. By comparing the results to \pp, modifications in jet evolution caused by the presence of a cold nuclear medium can be studied.
}
\maketitle
\vspace{-1em}
\section{Introduction}
\label{intro}
In these proceedings, three measurements performed by the ALICE Collaboration % (associated with \href{https://indico.cern.ch/event/1339555/contributions/6040810/}{this talk} at Hard Probes 2024)
of a novel jet substructure observable called the energy-energy correlator (EEC) \cite{Chen_2020,Lee_2025,PhysRevLett.130.051901} are presented. The EEC captures the correlation function of energy flow inside jets. It is experimentally defined as an energy-weighted two-particle correlation between pairs of particles inside a jet, as a function of the angular distance (\rl) separating the pair. The energy-energy correlation function, $\Sigma_{\rm EEC}(\rl)$, is defined as follows:
\begin{equation}
    \Sigma_{\rm EEC}(\rl) = \frac{1}{N_{\rm jet}\cdot\Delta} \int_{\rl-\frac{1}{2}\Delta}^{\rl+\frac{1}{2}\Delta}\sum_{\rm jets}\sum_{i,j}\frac{p_{{\rm T},i}p_{{\rm T},j}}{(\pt^{\rm ch\> jet})^2}\delta(\rl'-R_{{\rm L},ij})d\rl'.
\end{equation}
Here, $\pt^{\rm ch\> jet}$ refers to the total charged-particle jet \pt. The sum runs over all particle pairs $(i, j)$ inside the jet. For each pair, an energy weight is calculated: $p_{{\rm T},i}p_{{\rm T},j}/(\jpt)^2$. The weighted number of track pairs is counted as a function of the angular distance between both tracks, $R_{{\rm L}, ij} = \sqrt{(\varphi_j-\varphi_i)^2 + (\eta_j-\eta_i)^2}$, where $\varphi$ and $\eta$ are the azimuthal angle and pseudorapidity respectively. The angular bin width is $\Delta$, and the EEC is normalized by the total number of jets $N_{\rm jet}$. 

The EEC distribution is characterized by a clear transition region. This is sensitive to the transition between the perturbative and non-perturbative regimes of jet evolution. As jets are angularly ordered objects, with earlier splittings being wider than subsequent ones, the angular scale of the EEC is sensitive to the time evolution of jet formation and allows the eventual confinement of partonic jet constituents to be probed.

\section{Final measurement of EECs in \pp collisions}
\begin{figure}[h]
\centering
\hspace{-1em}
\vspace{-1em}
\begin{minipage}{.51\textwidth}
  \centering
  \includegraphics[width=\linewidth]{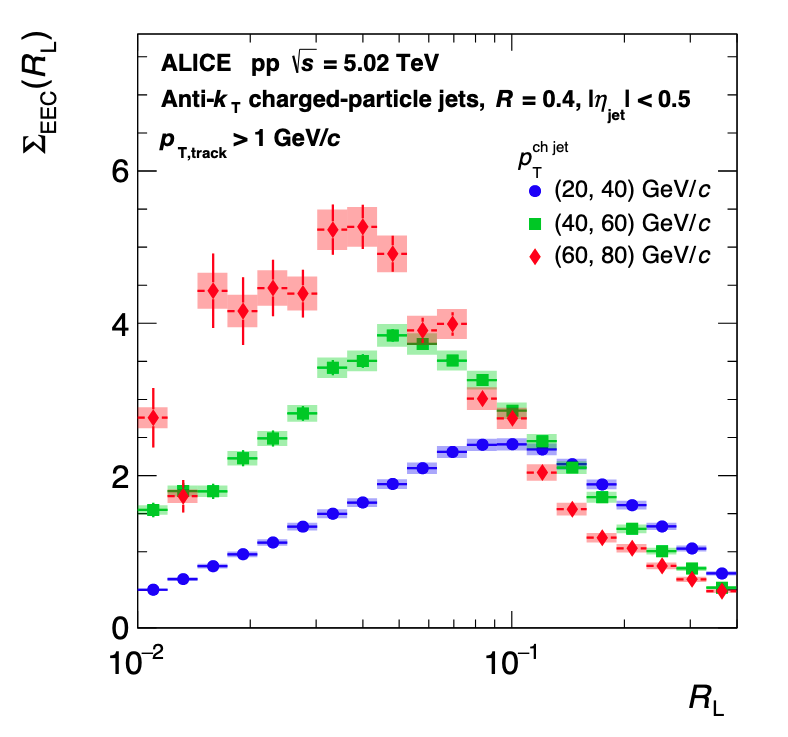}
  \vspace{0.2em}
  \caption{EEC in \pp collisions in three \jpt ranges: 20–40, 40–60, and 60–80 GeV/$c$ \cite{wfEECpp}. A clear \jpt dependence is seen in the EEC peak, with the distribution shifting to smaller angles as \jpt increases.}
  \label{fig:pp1}
\end{minipage}%
$\quad$
\begin{minipage}{.4\textwidth}
  \centering
  \includegraphics[width=\linewidth]{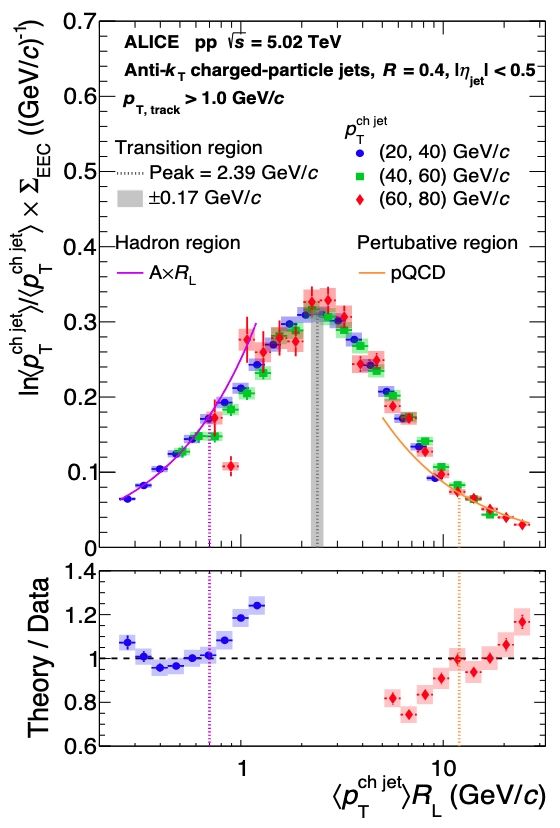}
  \vspace{-2em} 
  \caption{EEC rescaled as a function of \ptrl, showing the universality of the EEC transition region \cite{wfEECpp}.}
  \label{fig:pp2}
\end{minipage}
\vspace{-1.5em}
\end{figure}

First, the finalized ALICE measurement of EECs in \pp collisions is presented \cite{wfEECpp}. Figure \ref{fig:pp1} shows the \pp EECs in three \jpt ranges. A clear \jpt dependence in the EEC peak position is observed, with the distribution shifting to smaller angles as \jpt increases. Figure \ref{fig:pp2} shows a universal shape that emerges when we the EEC $x$-axis is rescaled from \rl to \ptrl, which is analogous to virtuality. The \jpt dependence disappears, and the EEC distributions collapse onto one curve with a clear peak position around 2.4 \GeVc. Figure \ref{fig:pp2} also shows a comparison to a pQCD calculation \cite{Lee_2025} -- good agreement is observed at very large \rl, suggesting that this region of the EEC maps onto perturbative physics. Similarly, at very small \rl, good agreement with the linear scaling for free hadrons is observed. Together this suggests that the EEC allows a separation of the perturbative and non-perturbative regions. It is worth noting that neither the pQCD calculation nor the linear scaling describe the EEC near the peak, suggesting that the transition between these regions is not sharp.

\section{First measurement of EECs in D$^0$-tagged jets in \pp collisions}
Having established a baseline measurement in inclusive \pp jets, EECs in heavy-flavor jets can now be studied. This has been studied in some depth theoretically \cite{hftheory}. Figure 1 in Ref. \cite{hftheory} shows a comparison between EECs in light-quark jets, charm jets, and beauty jets in PYTHIA. A clear mass dependence is observed; the charm and beauty EECs show strong suppression at small angles (a signature of the dead cone). The EEC offers a unique opportunity to probe the dead cone, as its amplitude quantifies the amount of radiation in a jet. 

\begin{figure}[h]
\centering
$\quad$\includegraphics[width=0.8\textwidth,clip]{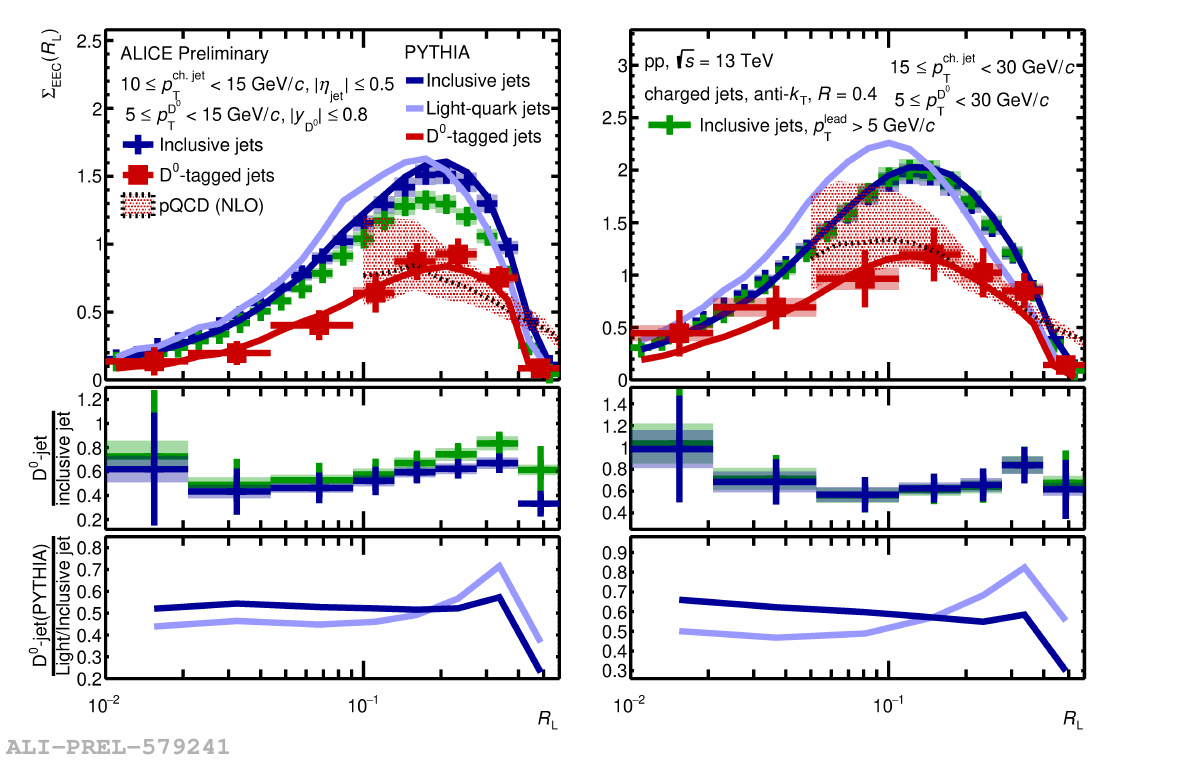} 
\caption{Measurement of EECs in D$^0$-tagged jets in \pp collisions at \s=13 \TeV. The left panel shows results for 10--15 \GeVc jets, and the right panels shows 15--30 \GeVc jets. The D$^0$-tagged jet EEC (red) is compared to inclusive jets (blue), PYTHIA (solid lines), and a pQCD calculation (shaded region).}
\label{fig:D0data}
\vspace{-1em}
\end{figure}

The ALICE measurement of D$^0$-tagged jets is shown in Fig. \ref{fig:D0data}. The measurement covers two \jpt ranges, 10--15 \GeVc and 15--30 \GeVc. Each jet is required to contain a D$^0$ meson with \pt $>$ 5 \GeVc. The upper panel compares the D$^0$-tagged jet EEC to inclusive and semi-inclusive EECs. The semi‑inclusive distribution applies a \pt selection on the leading track to study the fragmentation bias introduced by the D$^0$ \pt selection. % This also biases the jet sample slightly towards quark jets (inclusive jets are predominantly gluon jets). 

A visible suppression of the D$^0$-tagged EEC curves is observed relative to the inclusive EEC over the entire \rl range. These observations are consistent with the expected suppression of radiation from massive quarks due to the dead-cone effect. 
% The similarity between the light-quark jet EEC and inclusive jet EEC in PYTHIA indicates that the Casimir factor (for quarks versus gluons) has a small impact in this \jpt range, supporting the idea that this suppression is a consequence of the dead cone. 
The selection bias from the minimum D$^0$ \pt requirement appears to be negligible, based on comparisons of inclusive jets with and without a \pt$^{\rm leading} > 5$ \GeVc selection. The peak positions of the charm-tagged and inclusive jet EECs are similar in these \jpt ranges, and PYTHIA reproduces this. This similarity occurs because inclusive jets are predominantly gluon-initiated, and Casimir effects counterbalance the dead-cone in terms of peak position. In contrast, the measured peak shows tension with the next-to-leading-order (NLO) pQCD calculation \cite{hftheory}, suggesting the impact of non-perturbative effects on the charm-quark shower and hadronization processes.

\section{Comparison of \pp EECs to event generator predictions}
Because the EEC has perturbative and non-perturbative regions, it is not unreasonable to associate the transition region with hadronization. This motivates comparisons of these measurements to models with different hadronization mechanisms. Four event generators are considered: PYTHIA 8 \cite{Skands_2014}, Herwig 7 \cite{Bellm_2016}, and Sherpa 2.2.15 with two different tunes (Lund and AHADIC) \cite{Bothmann_2019}. PYTHIA and Sherpa Lund both use Lund string models, and Herwig and Sherpa AHADIC use cluster models. 

\begin{figure}[h]
\centering
\includegraphics[width=0.78\textwidth,clip]{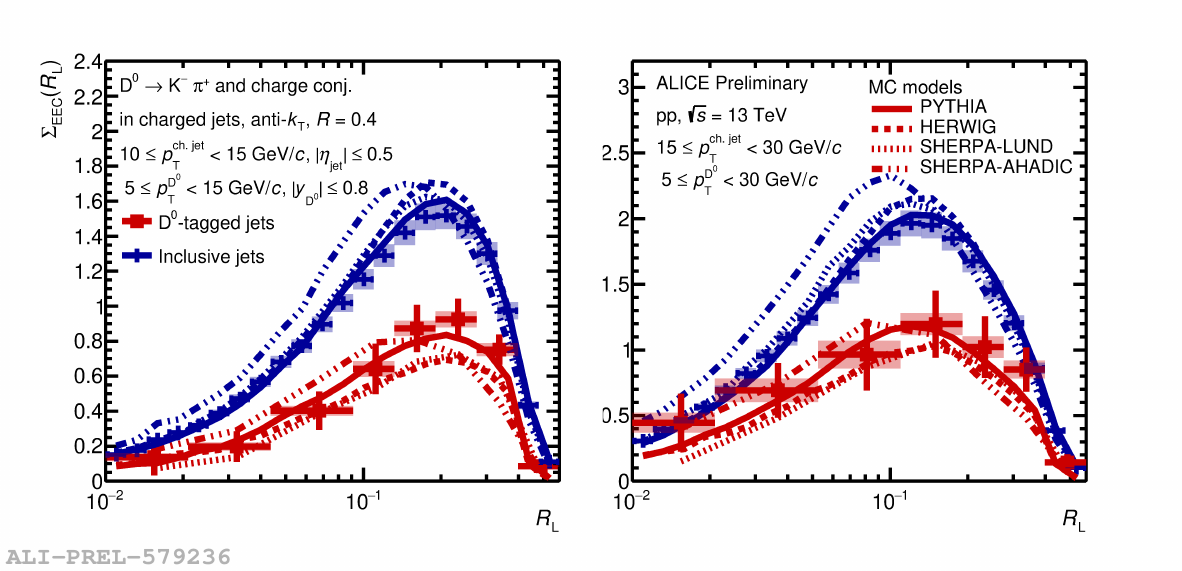} 
\caption{EECs as a function of \rl in D$^0$-tagged jets in data and four event generators. The left panel shows results for 10--15 \GeVc jets, and the right panels shows 15--30 \GeVc jets. The D$^0$-tagged jet EEC (red) is compared to inclusive jets (blue). }
\label{fig:ppmc}
\vspace{-1.5em}
\end{figure}

Figure 4 in Ref.~\cite{wfEECpp} shows a comparison of the inclusive \pp EEC measurement to these generators. The overall shape and amplitude of the EEC are sensitive to the hadronization mechanism, both at small and large \rl. However, the inclusive measurement does not clearly favor any particular hadronization model at the current level of precision \cite{wfEECpp}. Figure \ref{fig:ppmc} shows the same comparison for D$^0$-tagged jets, where PYTHIA is slightly favored. Herwig overpredicts the yield for inclusive jet EECs, but underpredicts heavy-flavor EECs. Sherpa Lund underpredicts the data everywhere, and AHADIC fails to describe the peak entirely. No definitive conclusions can be drawn from these comparisons. The inclusive measurements might suggest that cluster hadronization works better for higher \pt jets, and the D$^0$-tagged jets show an apparent preference for string-breaking models. However, neither statement is supported with overwhelming confidence by the current data.

\section{First measurement of EECs in inclusive jets in \pPb collisions}
% Now we turn our attention to \pPb collisions. These collisions are of course different from \pp events: the initial state is modified due to the different parton distribution function (PDF) of the Pb nucleus with resect to the proton. There is also the possibility of final-state effects, i.e. any effects occurring after the initial hard scattering. Studying EECs in \pPb gives us a window into the interactions at play in small systems and in cold nuclear matter. 
Measurements in \pPb collisions differ from \pp collisions because the parton distribution functions of the Pb nucleus differ from those of the proton, and final‑state effects may also be present. Studying EECs in \pPb collisions provides a window into interactions in small systems and cold nuclear matter.

\begin{figure}[h]
\vspace{-1em}
\centering
\hspace{-1em}
\begin{minipage}{.48\textwidth}
  \centering
  \includegraphics[width=\linewidth]{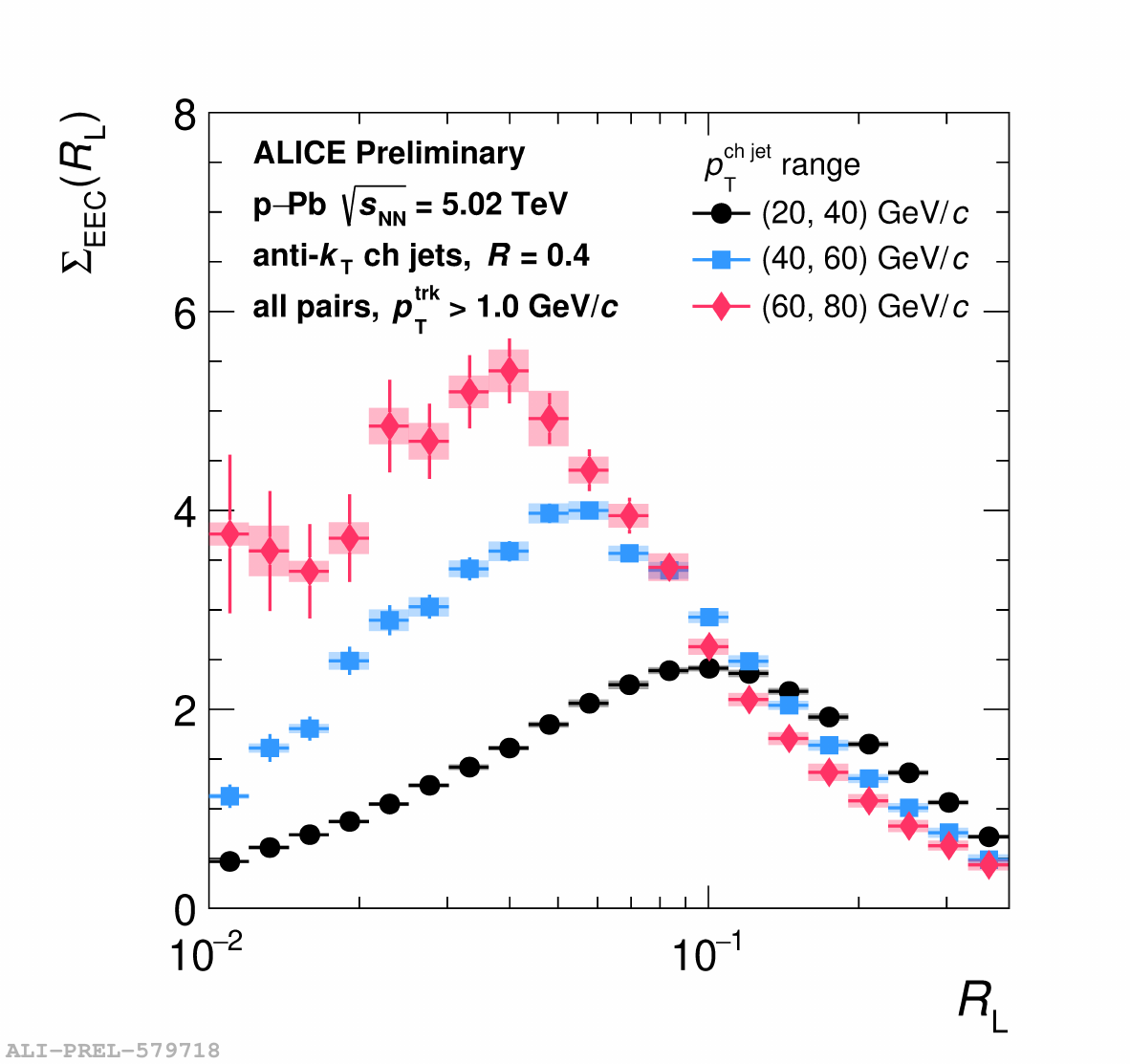}
  \caption{EECs for jets in \pPb collisions in three \jpt intervals.}
  \label{fig:ppb}
\end{minipage}%
$\quad$
\begin{minipage}{.48\textwidth}
  \centering
  \includegraphics[width=\linewidth]{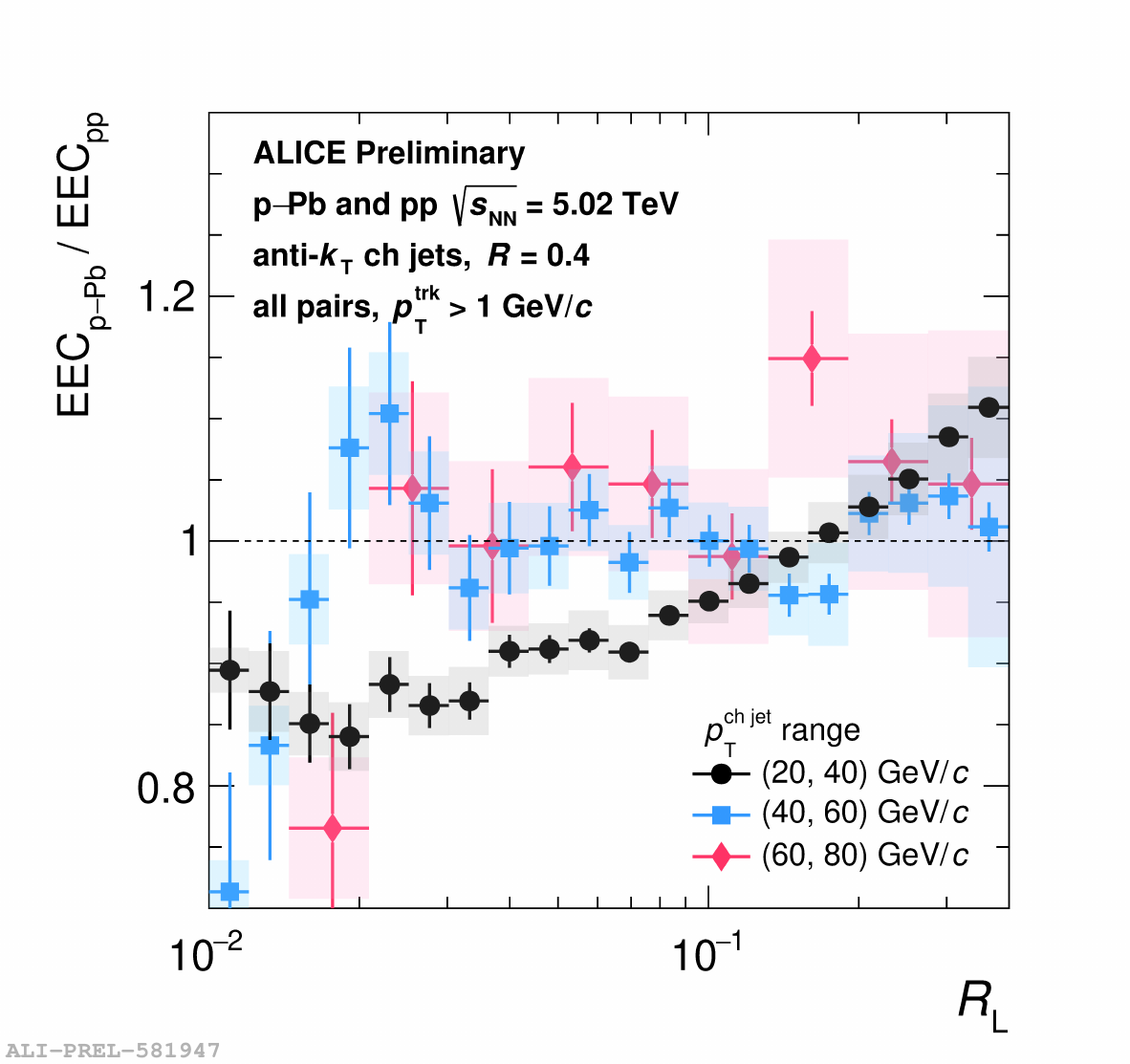}
  \caption{The ratio of the \EEC distributions in \pPb/\pp collisions in three \jpt intervals.}
  \label{fig:ppratio}
\end{minipage}
\vspace{-1.5em}
\end{figure}

EECs in inclusive jets in three \jpt ranges are considered, shown in Fig. \ref{fig:ppb}. This analysis requires a background subtraction procedure with two steps. First, the \jpt is corrected for the underlying event (UE) following the procedure in Ref.~\cite{rhosub}. The second step is a correction to the combinatorial background in the EEC distribution. Some of the pairs in the EEC distribution will have one or more tracks from the UE. A perpendicular cone method is used to estimate and subtract the contribution from these contaminating pairs. The same procedure is applied to the \pp baseline. After corrections, the ratios of the \pPb EEC distributions to the \pp EEC are shown in Fig. \ref{fig:ppratio}. In the lowest \jpt range, a significant difference between EECs in \pPb and \pp collisions is observed. There is a distinct enhancement at large \rl, coupled with a suppression at small \rl. This modification is not visible at higher \jpt. 

\begin{figure}[h]
\centering
\hspace{-1em}
\includegraphics[width=0.5\textwidth,clip]{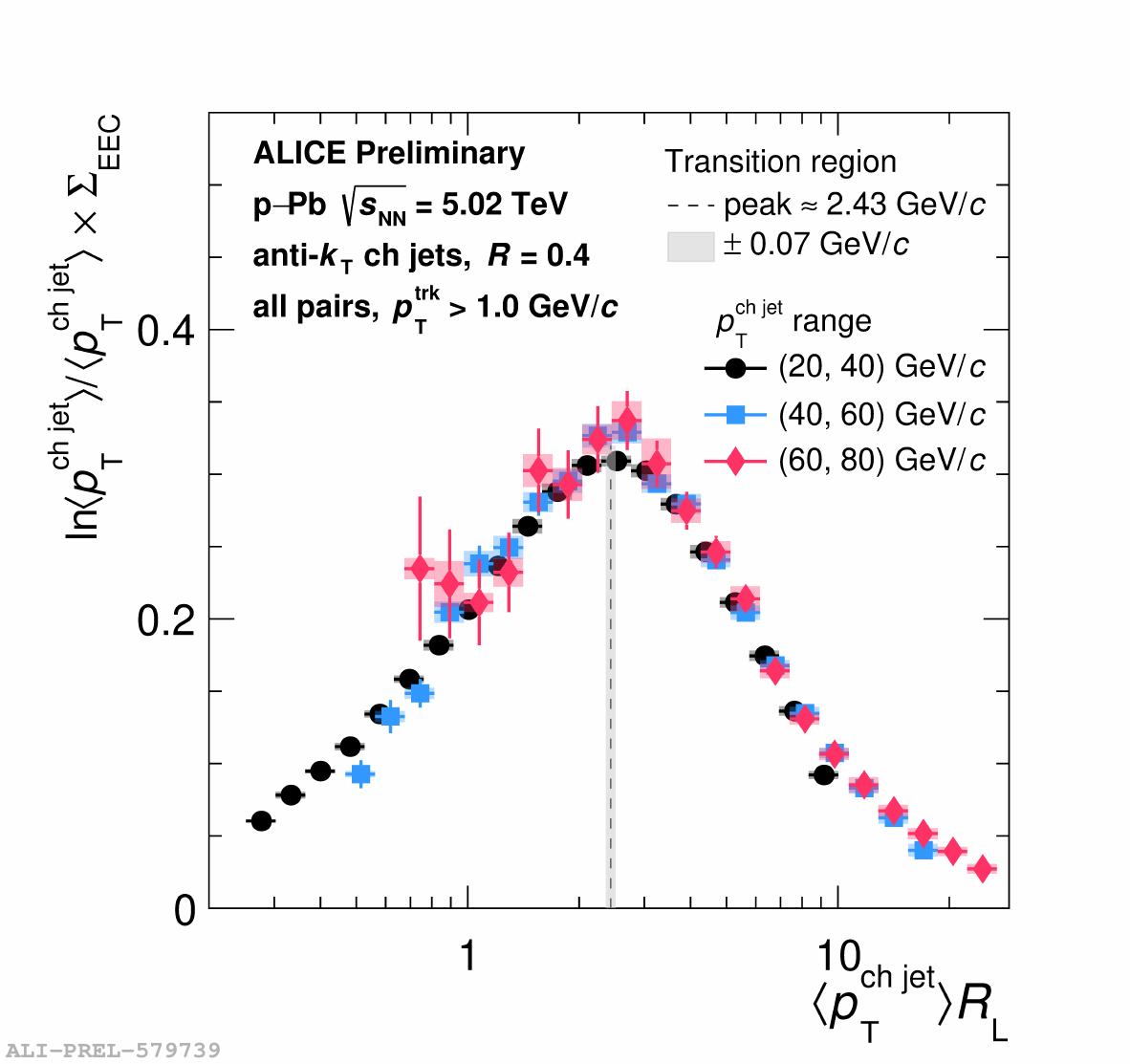} 
{\centering \caption{The \pPb EECs rescaled as a function of \ptrl, in three \jpt intervals. }}
\label{fig:ppbuniversal}
\vspace{-1.5em}
\end{figure}

Next, the \pPb data are inspected for the universal transition behavior found in \pp. After performing the same \ptrl rescalings, Fig. \ref{fig:ppbuniversal} shows that the \pPb jets collapse onto a common shape as well. Strikingly, the peak position (2.4 \GeVc) and peak height (0.32 \GeVc$^{-1}$) in \pPb are consistent with the values extracted in \pp, suggesting a universality across these systems. 

In all the plots presented so far, a \pt selection of 1 \GeVc is applied on the tracks from which pairs are built. Varying this cut (lowering it to 150 \MeVc and raising it to 2 \GeVc) allows its impact on the EEC to be studied. In Fig. \ref{fig:trkpy}, the track cut has little impact on the \pPb/\pp ratio at higher \jpt, but changing the threshold modifies the enhancement at large \rl in 20-40 \GeVc jets while leaving the low-\rl suppression unchanged. 

\begin{figure}[!h]
\centering{}
\includegraphics[scale=0.21]{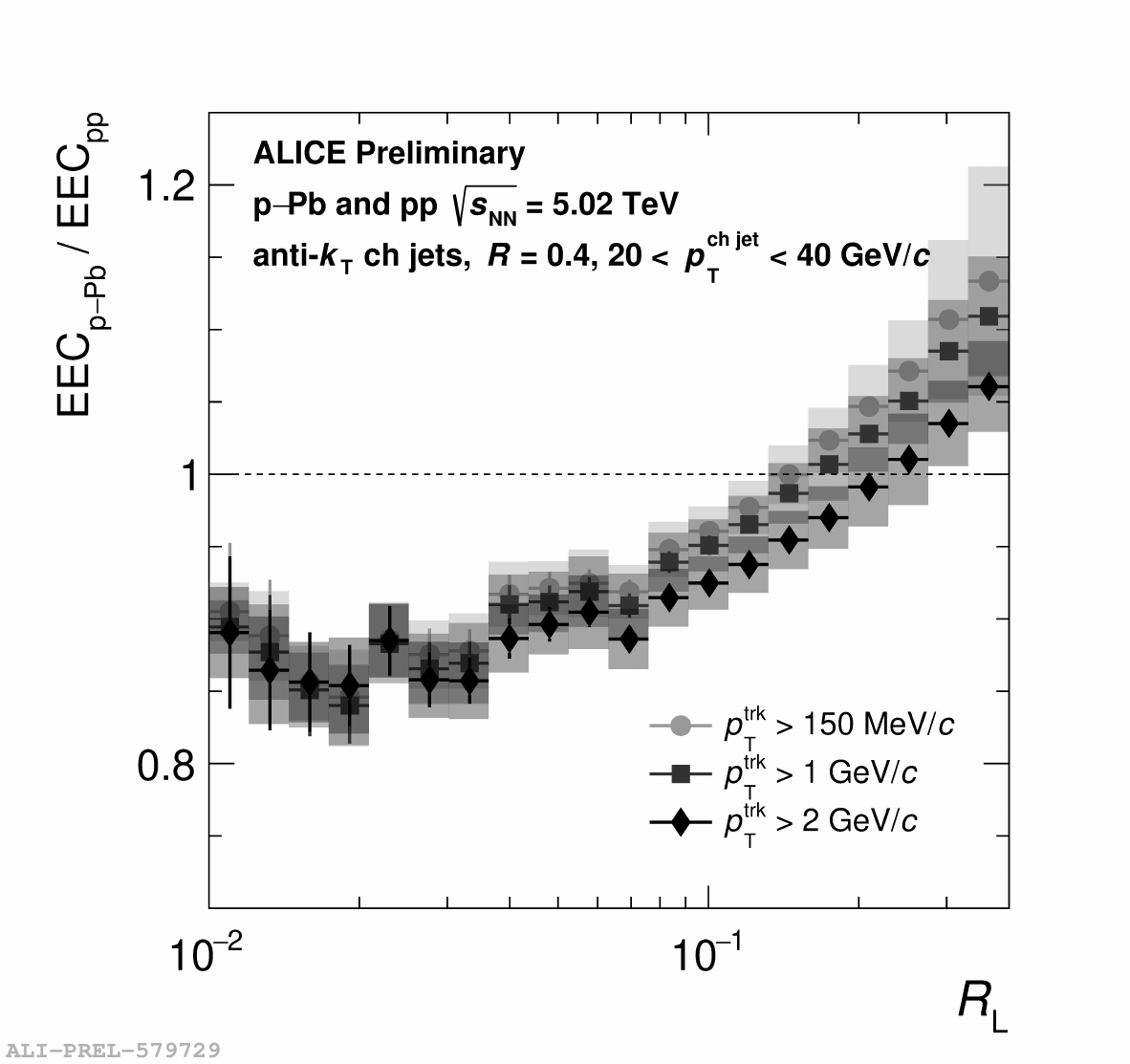} $\!$
\includegraphics[scale=0.21]{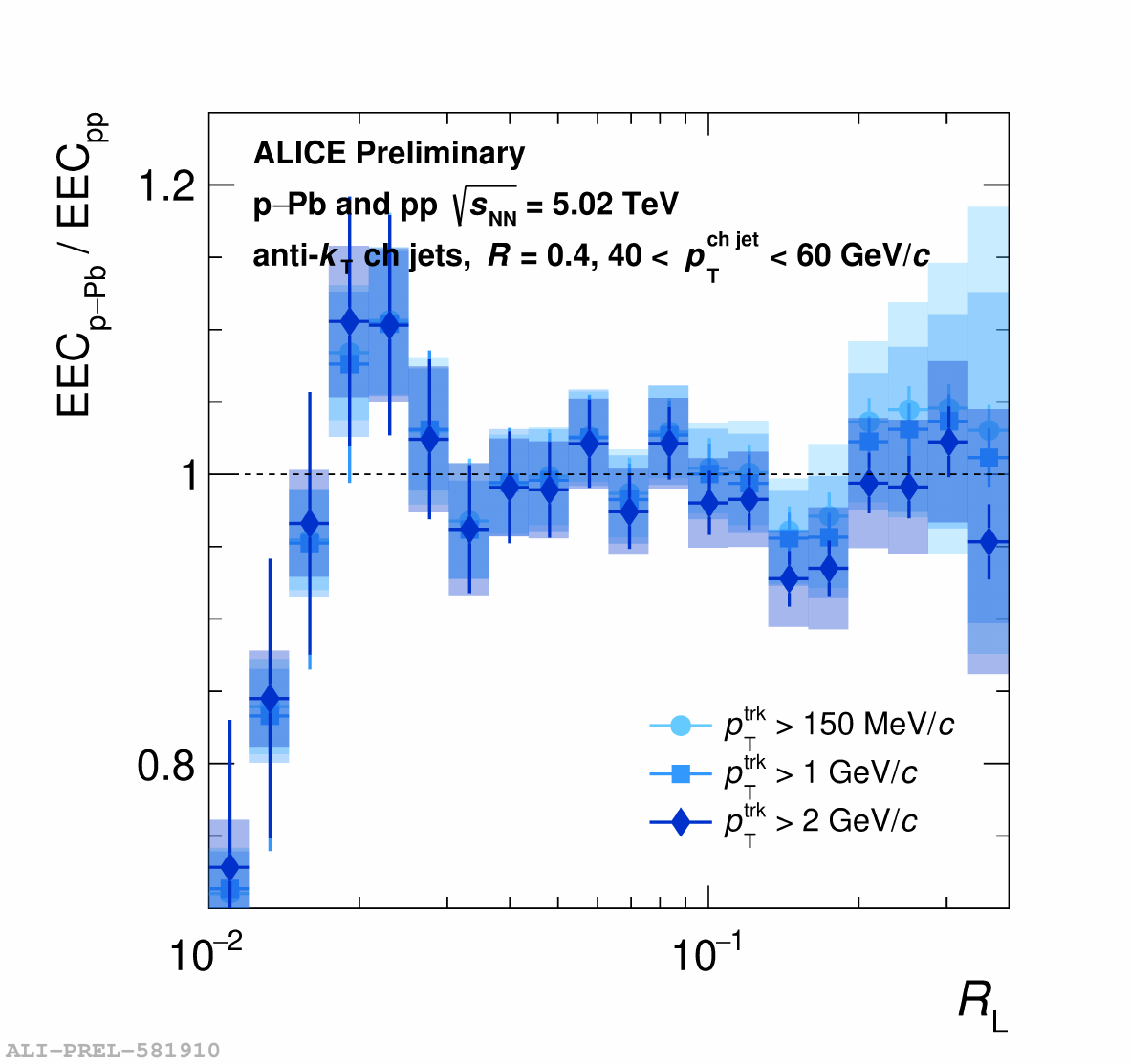} $\!$
\includegraphics[scale=0.21]{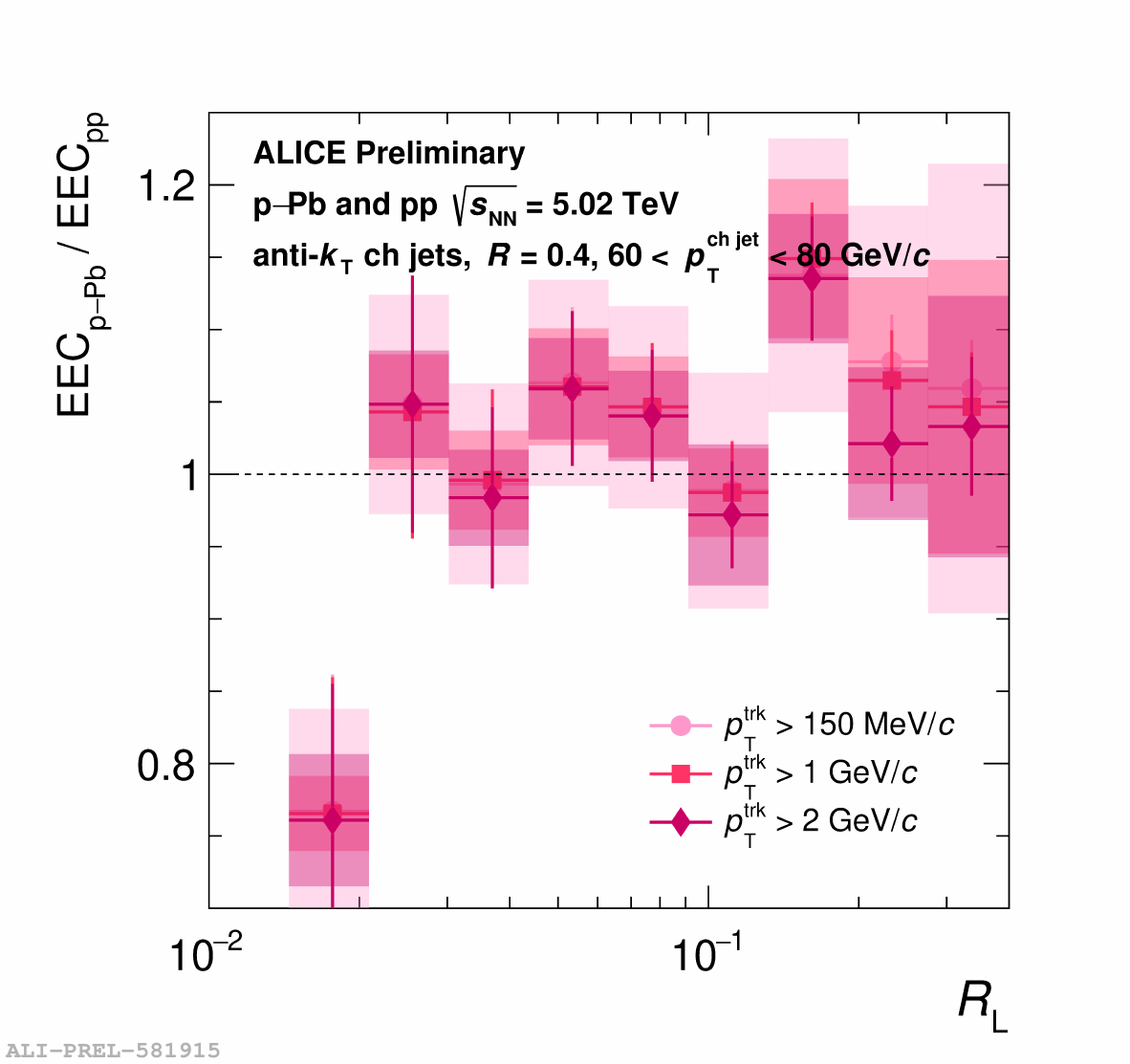}
\caption{The \pPb{}/\pp{} ratios of EECs with varying track \pt threshold. }
\label{fig:trkpy}
\vspace{-2em}
\end{figure}

Finally, Fig. \ref{fig:trkpy} compares the data to theoretical models. On the left is a comparison to a CGC model with varying saturation scales. This purely initial-state model fails to capture the data. In the middle are models in PYTHIA and PYTHIA Angantyr with nuclear PDFs. Standard PYTHIA shows a flat \pPb/\pp ratio. Angantyr, which models the Pb nucleus more comprehensively, replicates the suppression at small angles but fails to capture the large-\rl enhancement. The right panel presents a higher‑twist calculation with varying path length $L$, showing that certain final‑state effects can qualitatively reproduce the trends in the data. This work has been refined recently; more details can be found in Ref.~\cite{berndt}.

\begin{figure}[!h]
\centering{}
\includegraphics[scale=0.21]{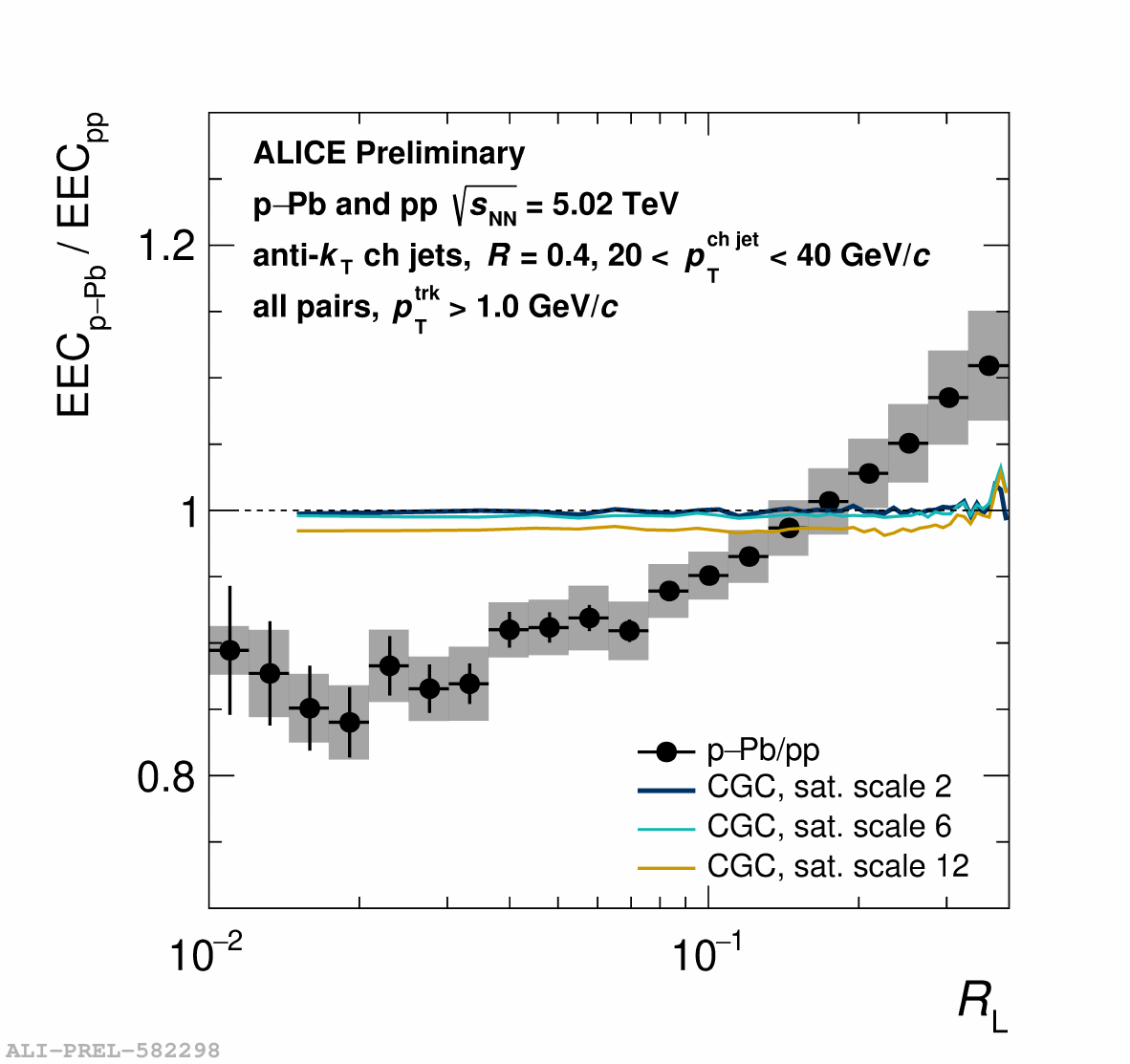} $\!$
\includegraphics[scale=0.21]{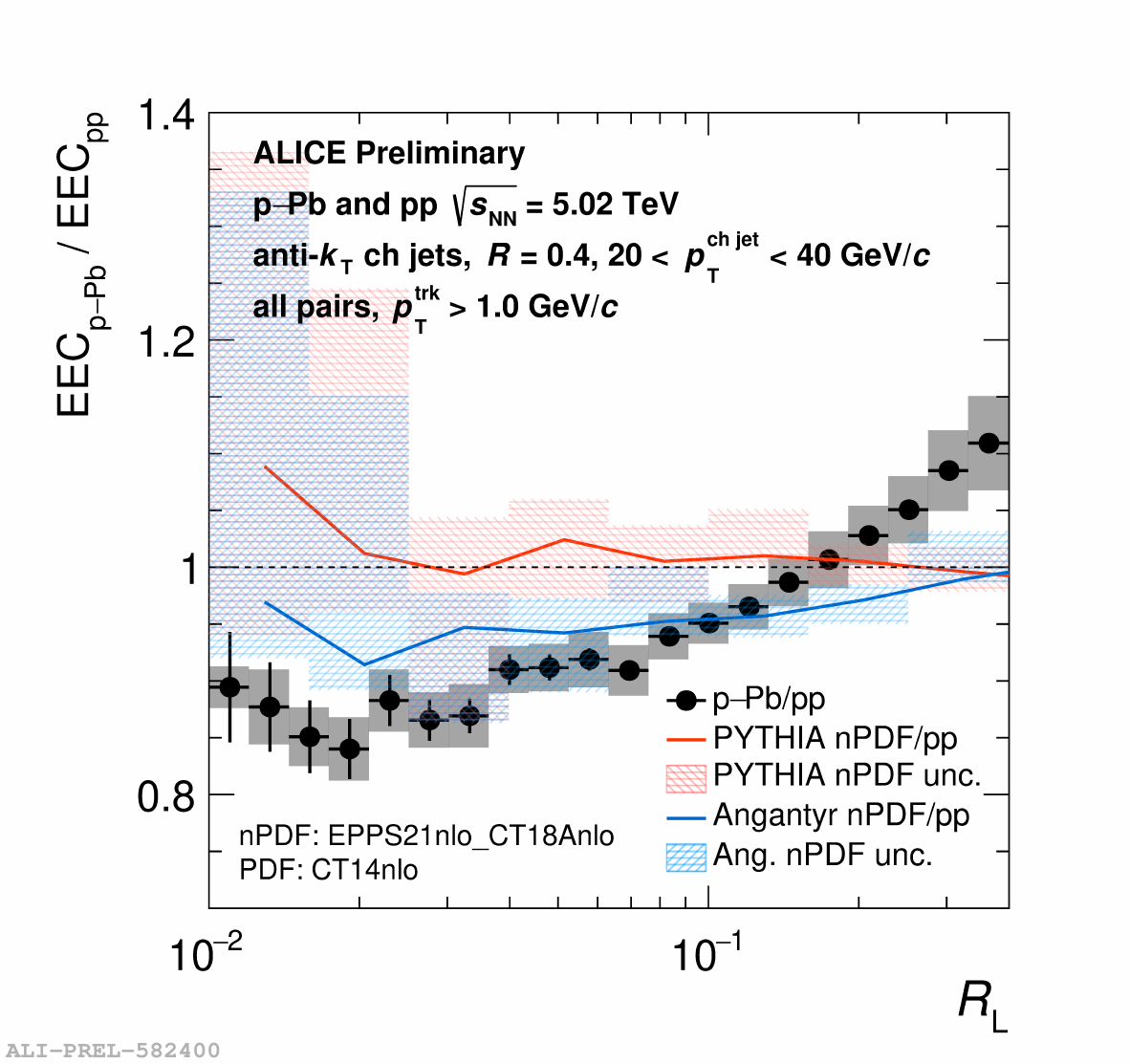} $\!$
\includegraphics[scale=0.21]{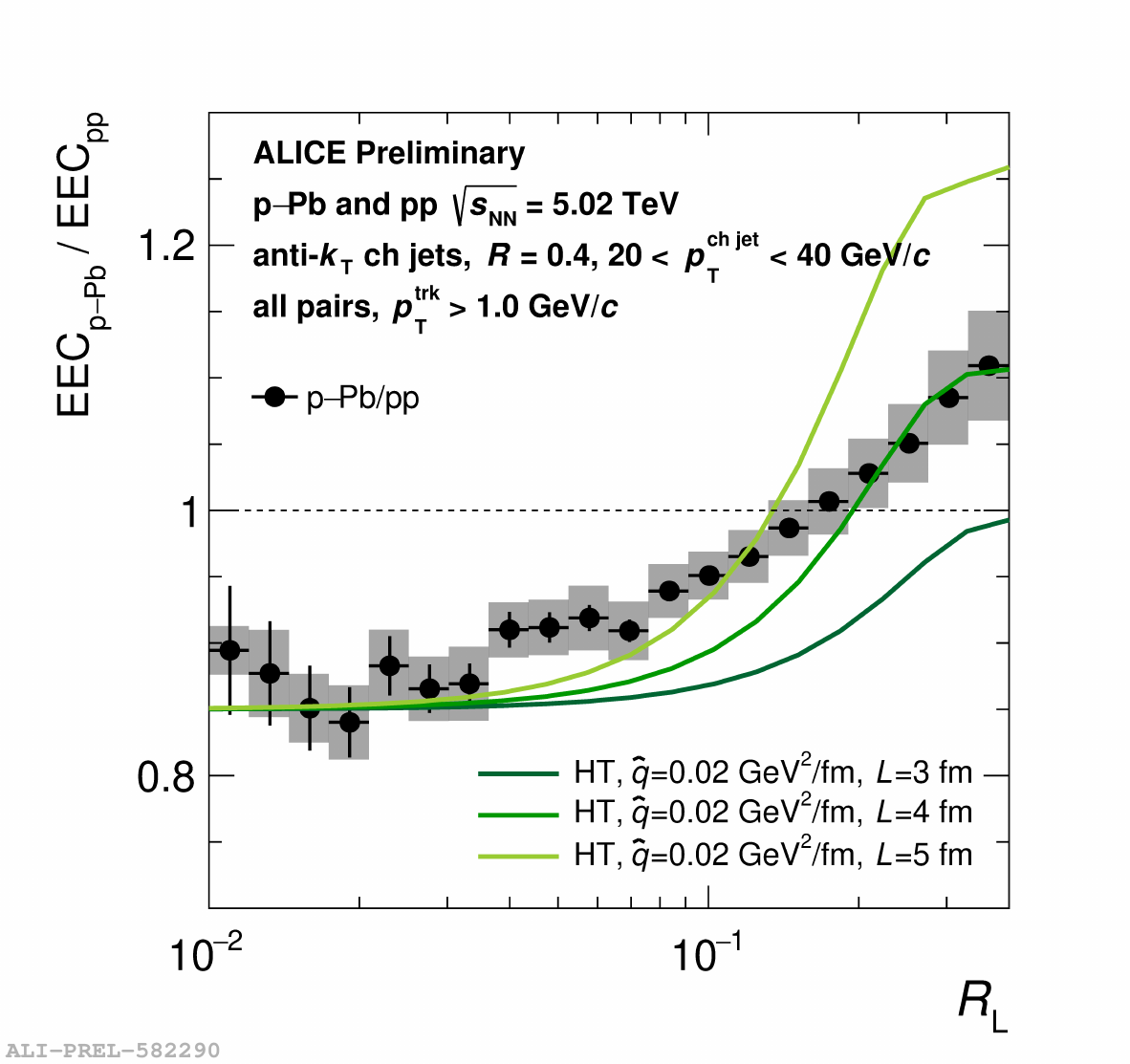}
\caption{Comparisons of the measured EEC distributions in 20--40 \GeVc jets to various calculations and models. From left to right: a CGC model, nPDF models in PYTHIA, and a higher-twist calculation.}
\label{fig:trkpy}
\vspace{-2em}
\end{figure}

In summary, three new EEC measurements from ALICE are presented. The inclusive \pp measurement revealed a universal EEC shape. The heavy-flavor measurement showed evidence of mass and flavor effects modifying charm jet EECs. The \pPb measurement shows a modification relative to \pp that cannot be explained by current initial-state models. 

\vspace{-1em}
\subsection*{Acknowledgments}
We gratefully acknowledge Haoyu Liu, Ian Moult, and Xiaohui Liu for providing CGC model predictions.

%
% BibTeX or Biber users please use (the style is already called in the class, ensure that the "woc.bst" style is in your local directory)
% \bibliography{your_bib_file} % Replace "your_bib_file" with the actual name of your .bib file
%
\vspace{-1em}
\bibliography{ref.bib}

\end{document}